# Environmental assessment of a new generation battery: The magnesium-sulfur system

Claudia Tomasini Montenegro[a], Jens F. Peters[b], Manuel Baumann[c], Zhirong Zhao-Karger[a], Christopher Wolter[d] and Marcel Weil*[a,c]

[a]Helmholtz Institute Ulm for Electrochemical Energy Storage (HIU), Ulm, Germany.
[b]University of Alcalá (UAH), Department of Economics, Alcalá de Henares, Madrid, Spain
[c]Institute for Technology Assessment and System Analysis (ITAS), Karlsruhe Institute of Technology (KIT), Karlsruhe, Germany.
[d] Custom Cells Itzehoe GmbH, Fraunhoferstraβe 1b, Itzehoe, Germany.

*Corresponding contributor. E-mail: marcel.weil@kit.edu

## Abstract

As the electrification of our economy and the corresponding increase in demand for battery storage systems are mostly driven by environmental concerns, information about the potential environmental impacts of the different battery systems is required. However, this kind of information is scarce for emerging post-lithium systems such as the magnesium-sulfur (MgS) battery. Therefore, we use life cycle assessment following a cradle-to-gate perspective to quantify the cumulative energy demand and potential environmental impacts per Wh of the storage capacity of a hypothetical MgS battery. Furthermore, we also estimate global warming potential, fossil depletion potential, ozone depletion potential and metal depletion potential, associated with the MgS battery production. The battery is modelled based on an existing prototype MgS pouch cell and hypothetically optimized according to the current state of the art in lithium-ion batteries (LIB), exploring future improvement potentials. It turns out that the initial (non-optimized) prototype cell cannot compete with current LIB in terms of energy density or environmental performance, mainly due to the high share of non-active components, decreasing its performance substantially. Therefore, if the assumed evolutions of the MgS cell composition are achieved to overcome current design hurdles and reach a comparable lifespan, efficiency, cost and safety levels to that of existing LIB; then the MgS battery has significant potential to outperform both existing LIB, and lithium-sulfur batteries.

**Keywords**: life cycle assessment, energy storage, post-lithium battery, lithium-ion battery and climate change

# 1 Introduction

The energy sector is and will remain a cornerstone of the social and economic development of society. Thus, the transition towards limiting the global temperature between 1.5 and a maximum of 2.0 degrees Celsius threshold, as defined in the Paris Agreement [1, 2], depends on the identification of pathways that contribute to the decarbonization process of this sector, which is the main contributor to global carbon emissions [1]. Consequently, the integration of renewable energy sources and the development of energy storage systems that enhance the flexibility of the power systems, while allowing them to buffer fluctuations associated with variable renewable energy, is a crucial factor to reach the energy-related sustainable development goals (i.e. SDG3, SDG7 and SDG13) [3, 4]. Despite the dominance of pumped hydro storage systems, batteries, and more specifically lithium-ion batteries (LIB), have become a crucial technology for this kind of service (contributing to 41% of the 0.34 GW non-pumped hydro storage) [5]. Also, there is a soaring demand for LIB for the transport sector, which faces similar challenges in terms of decarbonization [6, 7].

The recent success of LIB in these sectors is explained by their excellent performance, namely their high energy density, high power rate capability, long life cycle and flexible scaling [6]. Nevertheless, especially for the transport sector, further improvements are required for them to become mass technology. The main challenges for current LIB in this regard are: (i) improved energy density (300 Wh/kg),  (ii) cost reductions (below 100 US/kWh) [4],  (iii) increased lifetime (10 years)  [8], (iv) reduction of the content of scarce and critical resources [9, 10], and (v) reduced risk of dendrite formation [11, 12]. In particular, questions about the environmental impacts, demand for critical and scarce resources, and recyclability have come to the centre of public debate recently [4]. These concerns  are becoming more and more relevant during the decision-making process about different energy storage options [13, 14]. It is recognized that to avoid environmental burden shifting, the estimation of these impacts should be conducted considering a holistic approach along the life cycle of the LIB [14, 15]. Currently, such evaluation is carried out using life cycle assessment (LCA), which is an environmental assessment tool that considers inputs and outputs to estimate the potential environmental impacts based on the life cycle of a product [16, 17].  In the European Union, a harmonized version of the life cycle approach has served as a basis of the Single Market Green Initiative, a unified set of rules that promotes understanding of the trade-offs, on an environmental basis, of commercially available products,  which includes  LIB rechargeable batteries [18].

To overcome the environmental constraints of current LIB, research on other battery chemistries is conducted to identify new technologies with competitive performance, but reduced environmental impacts and material requirements [5, 12]. Regarding the latter, the elimination of cobalt and nickel, but also copper and lithium, are within the scope, all associated with different concerns about their future availability and high environmental impacts during the mining phase [9]. Among the prospective potential candidates, magnesium-sulfur (MgS) batteries are considered as one of the most promising options to manage these concerns regarding safety, energy density, environmental impacts and resource availability associated with LIB [19, 20]. Concerning energy density,  the advantage of the MgS cell is explained by the divalent $Mg^{+2}$ nature, theoretically offering almost double the volumetric capacity compared to lithium [19, 21]. Furthermore, MgS cells are also safer due to the potential absence of dendrite formation for magnesium [21]. They are expected to have significant potential for cost reduction, relying on widely available and cheap materials like metallic magnesium for the anode manufacture [22, 23]. Thus, when combined with a sulfur cathode, the MgS electrochemical couple is a promising, high capacity, low cost and safe post-lithium option [23-25].

While the environmental impacts of LIB have been assessed by numerous works and with different scopes (see the comprehensive reviews presented by Sullivan *et al.* [26, 27] and Peters *et al.* [28]), this information is scarce for post-LIB options, with only a handful of assessments of sodium ion, lithium-sulfur, lithium-air, composite cathode, and advanced LIB published recently [28-33]. For MgS, the only existing environmental assessment is based on a laboratory pouch cell in a configuration that does not allow any comparison with current LIB [34]. Therefore, this work aims to fill this gap by conducting an LCA of a theoretical commercial MgS battery according to the ISO standards 14040/44 [16, 17], establishing the following objectives:

- (i) development of the life cycle inventory for the MgS cell and battery;
- (ii) identification of the main hot spots associated not just with global warming, but also other environmental impacts that include fossil, metal and ozone depletion potential;
- (iii) quantification of the cumulative energy demand (CED) associated with the life cycle of the MgS battery;
- (iv) provide a theoretical horizon of the environmental performance of the theoretical MgS battery analyzed in this work.

The results are compared with current lithium-based options, more specifically, a lithium-iron-phosphate (LFP) [33, 35], and a lithium nickel-manganese-cobalt (NMC) battery [35, 36]. Furthermore, a lithium-sulfur (LiS) battery is also considered as a benchmark for a possible evolutionary stage after current LIB [30]. This information will set the scene for the future assessment of MgS batteries, which might be required to quantify the associated environmental impacts and contribute to establishing guidelines for optimizing upcoming MgS prototypes under eco-design aspects [37].

## 2 Methods

### Goal and scope definition

The goal of this study is twofold. Firstly, it is aimed at the identification and quantification of the environmental impacts associated with a theoretical MgS battery based on the MgS cell developed by Wagner *et al.* [38]. Secondly, the hot spots of the estimated environmental impacts will also be identified. The scope of the LCA of the MgS battery is from cradle to gate, considering 1 Wh of energy storage capacity provided by the battery on a battery pack level as the functional unit (the unit of reference for estimating and comparing potential environmental impacts). The following impact categories are calculated based on the mid-point hierarchic perspective of the ReCiPe 2008 impact assessment method [39], using Open LCA version 1.9 [40]: (i) global warming potential (GWP), i.e., climate change, being the most present environmental concern in society currently; (ii) ozone depletion potential (ODP), i.e., the impact on the decomposition of the ozone layer ('ozone hole'), (iii) metal depletion potential (MDP), i.e., the depletion of abiotic mineral resources, being the use of scarce metals of particular concern for lithium batteries and one of the arguments used for promoting alternative battery chemistries [9], and (iv) fossil depletion potential (FDP), since despite current efforts to shift toward cleaner energy resources, fossil energy still contributes 40% to the European electricity mix [41]. Additionally, the cumulative energy demand (CED) is calculated as an indicator of the total energy investment (renewable and non-renewable) required for the manufacture of the battery, including all upstream processes and inputs [42, 43]. It is considered as a useful summary ('footprint') indicator during the decision-making process related to the sustainability assessment of batteries, but also energy technology in general [44-46].

### Data sources and assumptions

The life cycle inventory (LCI) of the MgS battery system under analysis is developed based on a hybrid approach. As a result, primary data obtained for an existing prototype pouch cell is used to model both cell components and assembly of the MgS cell [38, 47]. Since the inventory data has been derived primarily from a prototype pouch cell, a hypothetical industrial-scale production needs to be assumed for a meaningful assessment and comparison with existing secondary batteries. Here, we rely on data available for current lithium battery options, being the principal comparable process steps of cell manufacture [31]. Both data sources and assumptions corresponding to each element of the inventory are given below. Information for the background system has been sourced from the ecoinvent database version 3.5 [48]. A detailed description of individual LCI is provided in the supplementary information.

### Battery cell configuration

The pilot-scale MgS cell layout described in Table 1 forms the basis of our battery model. The cell is composed of an Mg foil anode combined with a sulfur cathode and an $Mg[B(hfip)_4)_2]$*DME (magnesium tetrakis hexafluoroisopropyloxy borate with dimethoxyethane as organic solvent) electrolyte, hereafter referred to as $Mg[B(hfip)_4]_2$ (0.3M) [38]. The magnesium foil serves both as anode and anode current collector, while the sulfur is coated onto an aluminium foil current collector with a 5% black carbon additive for increased conductivity, and 5% carboxy-methyl-cellulose/styrene-butadiene rubber (CMC/SBR) polymer binder. Cathode and anode are separated by a polyolefin membrane, which makes up 10% of the total cell mass in the prototype. The whole-cell is assembled in a simple (and comparably thick) pouch housing. Due to the preliminary stage of the development of this cell, information about its performance in terms of cycle life and efficiency is not available. However, information about the energy density of this early version prototype cell is available, amounting to 57 Wh/kg on pouch cell level.

For this reason, we rely on a theoretical battery model and optimize the layout of the MgS cell assuming a hypothetical future commercial cell, allowing the evaluation of its potential environmental impacts in comparison with different battery types. However, due to the specific challenges associated with the electrochemistry of the MgS system [21, 49, 50], we only modify the passive components. Consequently, the cell is optimized regarding the share of the pouch cell housing and separator, using average values achieved by current LIB as targets. These components are selected because based on their high contribution to the mass share in the prototype battery cell, they are assumed to be subject to optimization once produced commercially. In particular, the following (hypothetical) targets are set: (i) reduced cell pouch mass share from 45% wt. (baseline battery configuration; MgS-BL) to 3% wt. (MgS-Evo1), and (ii) reduce the thickness and correspondingly the mass share of the separator from 10% wt. to 2% wt. (MgS-Evo2), corresponding to the current state of the art in LIB [51]. The composition of each version of the MgS cell is given in Table 1.

Due to the early stage in the research and development of the MgS battery, no information about its potential commercial applications is available. Thus, it is assumed that the MgS cells are mounted in an automotive battery pack with a configuration similar to that of current LIB packs, considering an average composition of 5.5% BMS, 14.5% casing and 80% cell [51]. Also, due to the absence of further information in this regard, the battery pack composition has been assumed equal for each MgS cell layout (MgS-BL MgS-Evo1 and MgS-Evo2). The mass share of cells is maintained constant for the different evolutions, assuming that increasing energy density of the cells leads to improved storage capacity of the battery pack while maintaining the same size and weight. Information about the estimated energy density on a cell and battery level is provided in Table 1. The procedure to calculate the composition of the optimized MgS cell Evo1 and Evo2 is detailed in the supplementary information.

Table 1 Composition of the original and optimized MgS cell. MgS-BL: Baseline MgS cell (prototype cell as provided by Wagner et al. *[38]*.). MgS-Evo1: First optimization with reduced cell package mass; MgS-Evo2: Second optimization with reduced separator thickness

| Item | Material | MgS-BL (mg) | (%wt.) | MgS-Evo1 (mg) | (%wt.) | MgS-Evo2 (mg) | (%wt.) |
|---|---|---|---|---|---|---|---|
| **Battery cells** | | | | | | | |
| Anode | Mg foil | 427 | 6.4 | 427 | 11.2 | 427 | 29.0 |
| Cathode | Sulfur | 421 | 6.3 | 421 | 11.1 | 421 | 28.7 |
| | Binder | 5 | 0.1 | 5 | 0.13 | 5 | 0.3 |
| | Carbon | 5 | 0.1 | 5 | 0.13 | 5 | 0.3 |
| | Al Collector foil | 88 | 1.3 | 88 | 2.3 | 88 | 6 |
| Separator | Polyolefin | 700 | 10.4 | 700 | 18.3 | 29 | 2 |
| Electrolyte | $Mg[B(hfip)_4]_2 \bullet DME$ | 2060 | 30.7 | 2060 | 53.9 | 451 | 30.7 |
| Housing | Al composite | 3000 | 44.7 | 115 | 3 | 44 | 3 |
| Total (cell) | | 6706 | 100 | 3821 | 100 | 1478 | 100 |
| Energy density (cell) | -- | 57 Wh/kg | -- | 100 Wh/kg | -- | 259 Wh/kg | -- |
| **Battery pack (1 kg)** | | | | | | | |
| Battery cells | | 0.8 kg | 80% | 0.8 kg | 80% | 0.8 kg | 80% |
| Pack housing | | 0.145 kg | 14.5% | 0.145 kg | 14.5% | 0.145 kg | 14.5% |
| BMS | | 0.055 kg | 5.5% | 0.055 kg | 5.5% | 0.055 kg | 5.5% |
| Total (pack) | | 0.20 kg | 20% | 0.20 kg | 20% | 0.20 kg | 20% |
| Energy density (pack) | | 46 Wh/kg | -- | 80 Wh/kg | -- | 207 Wh/kg | -- |

### System and system boundaries

As shown in Figure 1, the system boundaries of the MgS product system under analysis includes the manufacturing of the cell components and battery assembly.

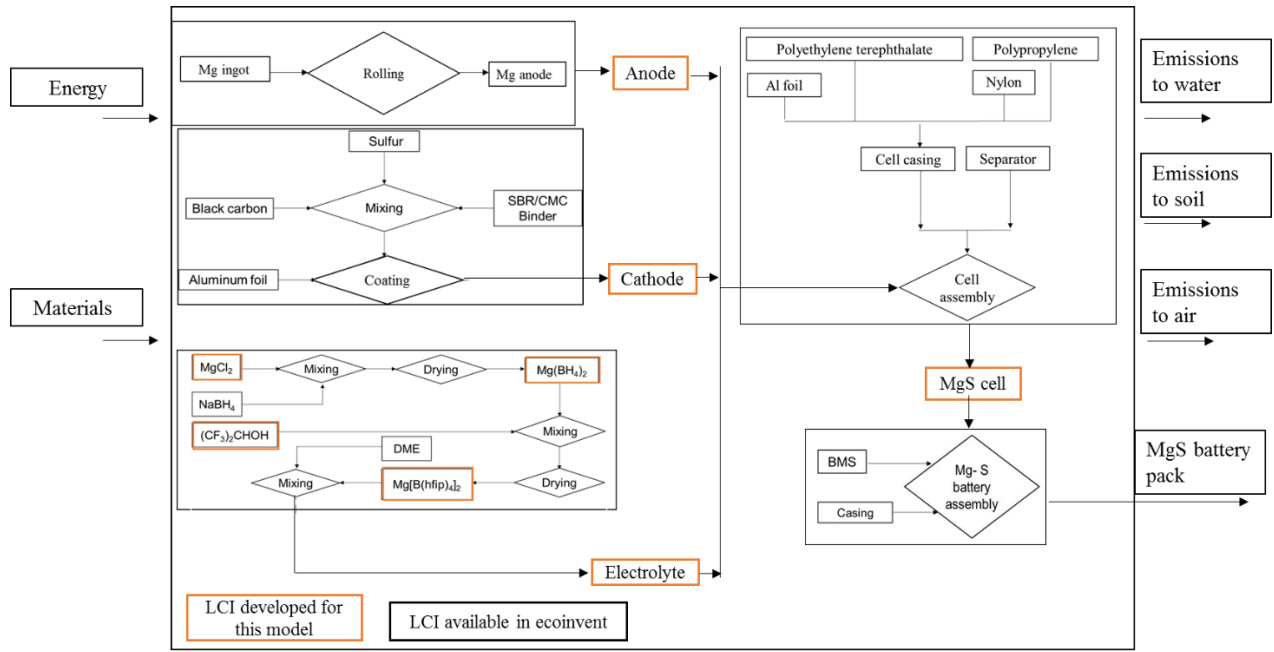

**Figure 1** System boundary of the MgS battery product system

Both use and recycling phases are disregarded due to insufficient knowledge about achievable cycle life and long-term stability. Before evaluating the MgS battery on an application level, the challenges associated with avoiding the formation of passivating layers need to be overcome [52]. Regarding the recycling phase, as pointed out by Mohr *et al*., available information about the environmental impacts of the recycling of lithium-ion batteries is scarce, and mainly based on unspecific data about the recycling process or the composition of the waste stream [53]. This stage should be carefully modelled to quantify the environmental impacts of the reuse of LIB [29, 54]. Therefore, due to the prevailing lack of data about the recycling of battery chemistries already consolidated in the market, information about the future candidates such as the MgS battery is either null, or not freely available. For this reason, this life cycle stage has not been considered within the system boundary of this analysis. This limits the study up to a certain point, but eliminates additional uncertainties from assumptions for these life cycle stages, and allows the estimation of targets in terms of cycle life that the MgS battery needs to achieve for competing with existing LIB.

### Life cycle inventory

MgS cell inventory

As seen in Figure 1, the LCI of the elements of the manufacture of the MgS cell encompasses data for the anode, cathode, electrolyte, cell housing and separator. Electricity supply required for both battery and cell manufacture has been modelled based on the European electricity mix. Due to the lack of information about anode manufacture, it is assumed that the production process of Mg foil is similar to that of other metal foils, like aluminium. For the rolling process, aluminium sheet rolling is considered as a valid proxy, as the rolling processes of Al and Mg show similar energy demands and associated $CO_2$ emissions [55]. In both cases, Mg ingot manufacture and the rolling process are sourced from ecoinvent 3.5 [48]. According to Wolter *et al.* [56], the thickness of 100μm for the anode foil is determined primarily by practical aspects, i.e., the commercial availability of foil with this thickness. Information for the background system has been sourced from the ecoinvent database version 3.5 [48]. A detailed description of individual LCI is provided in the supplementary information.

For proprietary reasons, the exact manufacturing process of the cathode is not disclosed, but its composition can be derived from the data available for the reference pilot MgS cell [38]. This process consists of a mixture of sulfur, black carbon and (CMC/SBR) polymer binder in mass ratios of (5:4:1). The cathode manufacturing process is modelled assuming a slurry-casting process similar to that of LIB, where the slurry mixture is cast on the aluminium current collector [57]. While background inventory data for electricity, heat and infrastructure, sulfur and black carbon have been sourced from ecoinvent 3.5 [48], the LCI of the CMR/SBR binder stems from Peters *et al.* [31]. Because of its electrochemical characteristics, in combination with its stability in the ambient atmosphere (air and water), $Mg[B(hfip)_4]_2$ (0.3M), is used as the electrolyte for the MgS cell [25]. The synthesis of the electrolyte salt $Mg[B(hfip)_4]_2$ is based on the dehydrogenation reaction of $Mg(BH_4)_2$ with $(CF_3)_2CHOH$ (hfip)[47], with the corresponding inventory data for hfip and its precursors estimated based on the hydrogenation of hexafluoroacetone [58].

Since no industrial-scale manufacturing process of $Mg[B(hfip)_4]_2$ is yet developed, both energy and infrastructure requirements have been approximated based on industrial sodium tetrafluoroborate production [59]. Since, as explained above, the manufacturing process of post-lithium batteries such as the MgS battery is assumed as similar to those of LIB, electricity and heat requirements for the cell and battery pack assembly processes are derived from averaged aggregated data from LCA studies on LIB production [51]. The components of the battery pack (housing, auxiliary components and battery management system) are modelled according to the current state of the art for LIB [36, 51].

# 3 Results

### Cumulative energy demand (CED)

Figure 2 provides the CED results for the three MgS cell configurations. For the baseline cell configuration (MgS-BL), the high contribution of the cell housing is evident, driven mainly by the aluminium content of the pouch and its energy-intensive production. Reducing the mass of the pouch housing (MgS-Evo1) not only reduces the contribution of the pouch housing, but also increases the energy density of the cell (less inactive material in the cell), and therefore reduces the impact of all other non-active components proportionally. The second optimization (MgS-Evo2) reduces the thickness of the separator, also affecting the amount of electrolyte, and thus further reducing total energy demand.

As seen in Figure 2, the CED associated with this cell amounts to 1583 Wh. This value can be considered as the total energy investment (including all upstream energy inputs along the process chain) and needs to be amortized during battery use. In this sense, it also gives a certain lower limit for the lifetime of the battery, since with lifetimes < 390 cycles the energy investment will always be higher than the return.

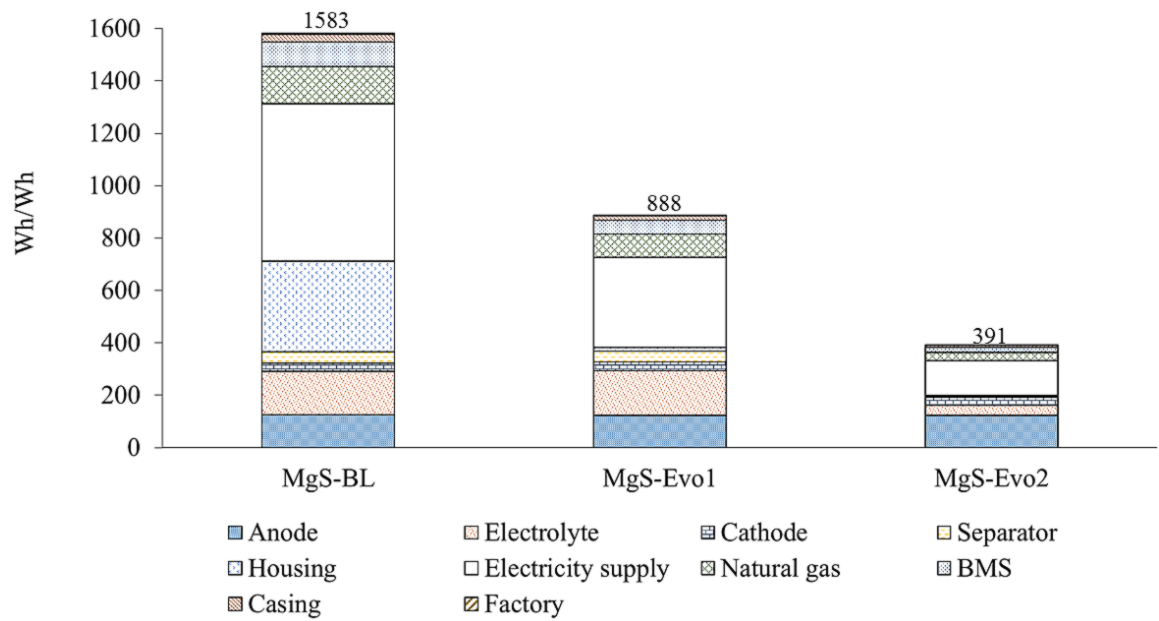

**Figure** 2 CED associated with the three MgS battery configurations. Legend (MgS-BL: baseline layout according to prototype cell; MgS-Evo1: first evolution with optimized pouch housing; MgS-Evo2: second evolution with optimized separator thickness and the correspondingly lower amount of electrolyte

### Environmental profile of the MgS-battery

A summary of the environmental profile of the MgS battery configurations is given in Figure 3.

Overall, the higher the energy density, the lower the environmental impacts. Similar to the CED results, the two evolutions of the baseline cell configuration (MgS-Evo1 and MgS-Evo2) show significantly lower environmental impacts than the baseline MgS-BL. A detailed analysis of the contribution of each component for the three different MgS battery configurations is given for each impact category in the following section.

Global warming potential

The contribution of the anode represents 41% and 20% of the total estimated GWP associated with MgS-Evo2 and MgS-Evo1, respectively. The majority of these greenhouse gas emissions can be explained by the electricity requirements of the Mg foil manufacture from the Pidgeon process (91%). The electricity requirements for the assembly of the cell and battery manufacture respectively make up another 25% and 32% of the total GWP of the MgS-Evo2 and MgS-Evo1 batteries. The electrolyte is identified as the third hot spot, respectively contributing 17% and 7% for the GWP associated with MgS-Evo1 and MgS-Evo2. This contribution is explained by the electricity requirements for the manufacture of the salt $(Mg[B(hfip)_4]_2)$ (39%).

Fossil depletion

As seen in Figure 3, anode manufacture can be identified as the central hot spot (39% of the total FDP), followed by the electricity required for the manufacturing process (mainly cell assembly) (27%) and electrolyte manufacture (10%). The main contribution of the FDP associated with the MgS-Evo1 battery comes from electricity supplied for cell and battery manufacture (27%). Additionally, the shares of the electrolyte (21%) and anode manufacture (18%) are identified as the second and third hot spots for this battery pack. About 61% of the electrolyte's share is caused by the manufacture of the salt (Mg[B(hfip)$_4$]$_2$, due to its electricity requirements (30%). The depletion of fossil resources associated with anode manufacture is mainly created by the magnesium foil manufacturing process (99%).

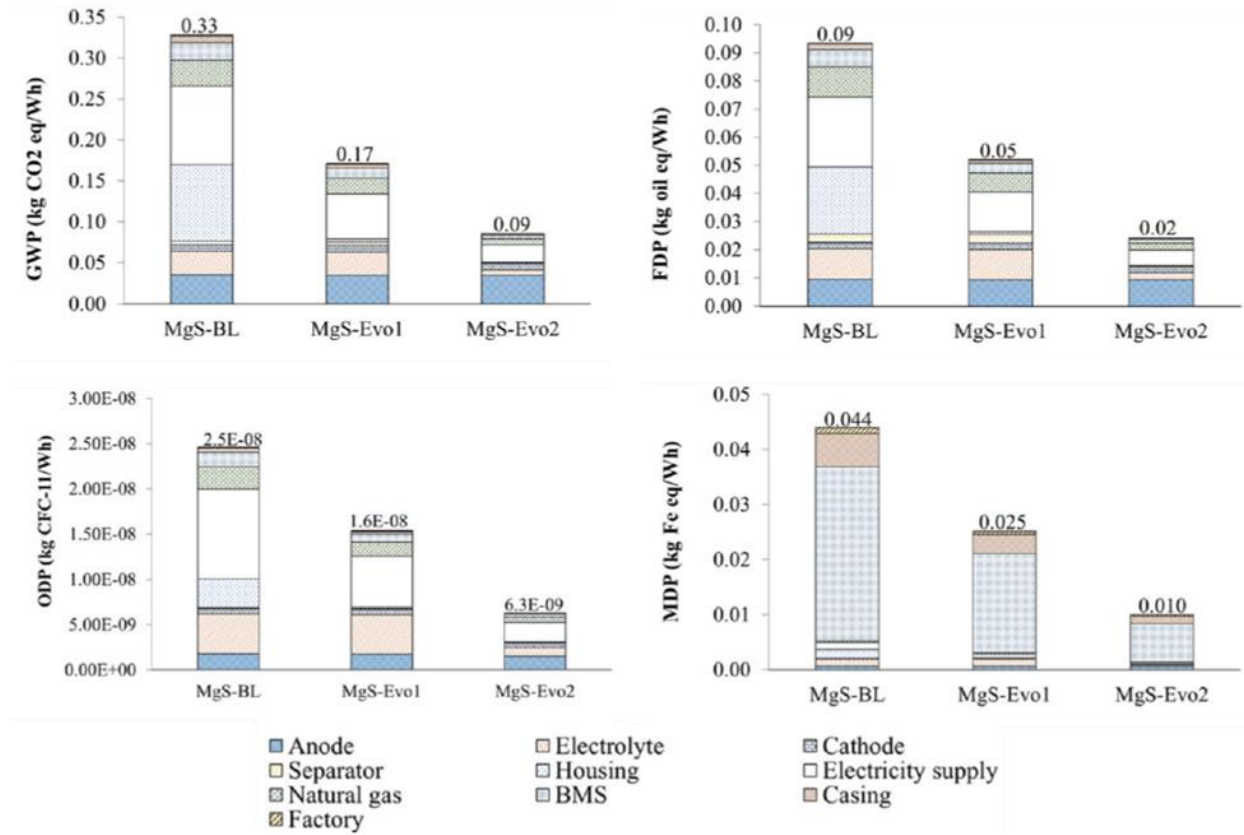

**Figure 3** Contribution analysis of the selected ReCiPe environmental impacts of the MgS-BL, MgS-Evo1 MgS-Evo2 batteries. Legend (MgS-BL: baseline layout according to prototype cell; MgS-Evo1: first evolution with optimised pouch housing; MgS-Evo2: second evolution with optimised separator thickness and the correspondingly lower amount of electrolyte.

Metal depletion potential

The contribution of the BMS represents the main hot spot (71%) of the estimated MDP associated with the MgS-Evo1 and MgS-Evo2, which is followed by the contribution from the casing of the battery pack (14%). Although information about the BMS and the casing of the MgS batteries is still missing, these results indicate that on a battery pack level, the components are critical drivers of the estimated MDP. As expected, the MgS battery cell itself hardly shows any MDP impacts, as it contains no scarce or critical materials.

Ozone depletion potential

Electricity supply for battery and cell manufacturing dominate the ODP of the MgS batteries, followed by the electrolyte for MgS-BL (18%) and MgS-Evo1 (29%), as well as the anode for MgS-Evo2 (24%); and thermal energy required for the manufacturing process. These decrease with increasing energy density and therefore, with the subsequent optimization stages.

## Comparison between the environmental impacts of the MgS battery and lithium batteries

To situate the results for the MgS battery within the current landscape, environmental impacts of the MgS battery are compared with three current technologies. These have been selected based on the maturity of the lithium battery market for automotive applications: 1) a nickel-manganese-cobalt lithium battery NMC, based on Ellingsen *et al*. (NMC (Ell)) and Majeau-Bettez *et al*. (NMC (M-B)) [35, 36]; 2) lithium-iron-phosphate (LFP), based on Zackrisson *et al*. (LFP (Zak)) and Majeau-Bettez *et al.* (LFP (M-B)) [33, 35], and 3) a theoretical lithium-sulfur (LiS (Deng)) battery considered to be a potential candidate as a future substitute for the LIB [30], based on a hybrid approach considering laboratory and pilot-scale information for the cell manufacture [30]. For these three systems, the specific layout and corresponding environmental impacts are estimated based on data from previous studies [30, 33, 35, 36], with the energy densities normalized by assuming identical cell housings and battery pack layout [51] (See Table 2).

**Table 2** Energy density, NMC and LiS battery systems

| Battery chemistry | Energy density Wh/kg (original value) | Energy density Wh/kg (adjusted value) |
|---|---|---|
| NMC (Ell)  [36] | 105.1 | 130.3 |
| LFP (Zak)  [33] | 93 | 86.4 |
| LiS (Deng)  [30] | 220 | 224 |
| NMC (M-B) [35] | 112 | 144.85 |
| LFP (M-B)  [35] | 88 | 113.8 |

The **GWP** results are displayed in Figure 4. The baseline MgS-BL shows by far the highest impacts, but reducing the mass of the pouch housing (MgS-Evo1) reduces the GWP impact to that of the LiS battery. However, further improvement is needed to outperform the LIB. Due to a reduction in the amount of electrolyte being soaked up by the separator, this improvement is achieved by reducing the thickness of the separator to values similar to that of current LIB, as modelled in the MgS-Evo2.

The $CO_2$eq. emissions from the cathode represent the main cause to the total GWP associated with the LFP (Zak), LFP (M-B) and NMC (M-B), contributing to 30.5% to the former and 41% to both LFP (M-B) and NMC (M-B). These contributions are explained by the emissions associated with tetrafluoroethylene manufacture. As seen in Figure 4, the electricity supply for the cell and battery assembly is the central hot spot for the other battery types, with a contribution between 25% for MgS-Evo2 up to 39% for the LiS. Still, the contribution to GWP of the remaining components is very different. For the MgS-Evo2 the second hot spot is the anode (40%), but for the LiS and the LIB, it is the cathode, contributing 25% and 36% to the total GWP, respectively.

Except for the LiS, **FDP** is dominated by the electricity demand for battery manufacture, 26% MgS-BL; 31% NMC (Ell); 30% (M-B)), followed by the cathode for the lithium-based batteries 20% (NMC (Ell); 21% NMC (M-B); 20% LFP (M-B); 43% LiS (Deng)), and the BMS for the LFP (Zak) (26%). The share of the electrolyte is also representative for magnesium-based batteries (21% MgS-Evo1; 10% MgS-Evo2), while for the baseline cell configuration MgS-BL, the cell pouch contributes the significant share with 26% of the total FDP.

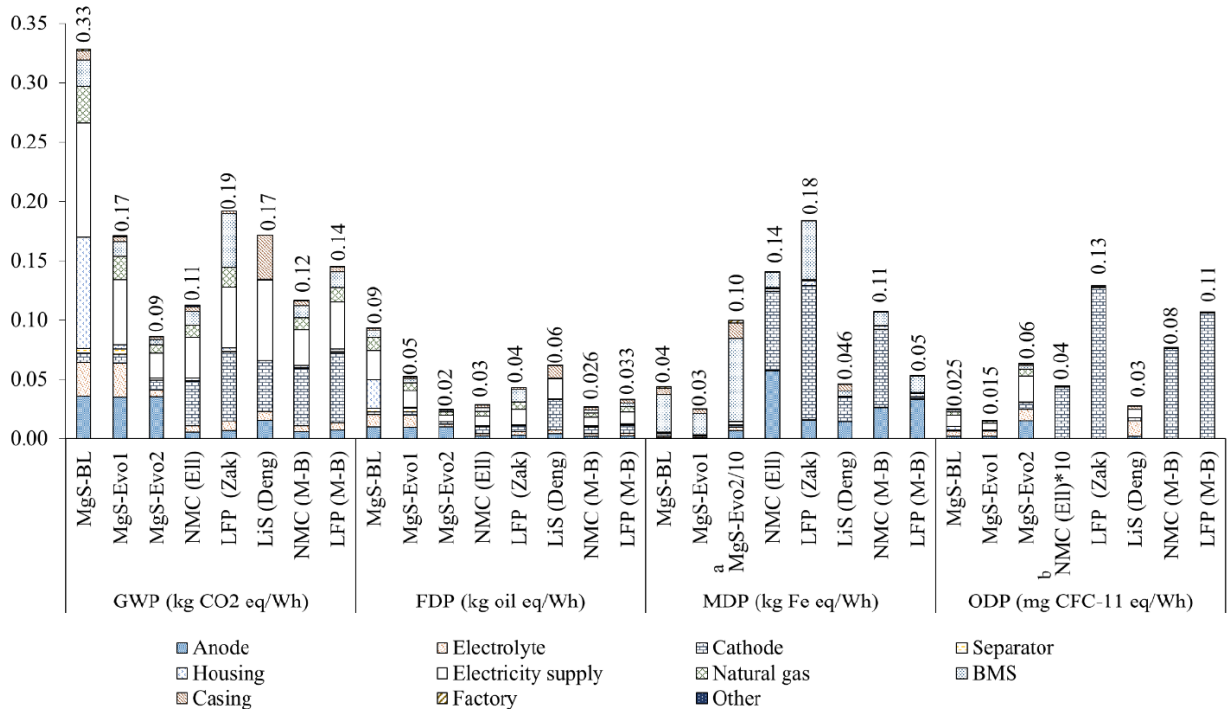

**Figure 4** Comparison between the environmental impacts associated with the MgS and lithium batteries

The **ODP** of the MgS batteries is mainly caused by the electricity demand for battery manufacture (MgS-BL: 40%; MgS-Evo1: 36%; MgS-Evo2: 34%), followed by the electrolyte (MgS-BL: 17.7%; MgS-Evo1: 28%; MgS-Evo2: 15.3%). For the NMC (Ell) 91%, NMC (M-B) 99%, and LFP (Zak) 98%, LFP (M-B) 99%, the main contribution is attributed to cathode manufacture. These contributions are explained by tetrafluoroethylene emissions to air from the binder manufacture (99%). While for the LiS (Deng), the electrolyte (47%) is the main hot spot for ODP, followed by the electricity supply (25%).

Under **MDP** aspects the BMS is the dominant contributor for all MgS batteries (72% MgS-BL; 71% MgS-Evo1; 69% MgS-Evo2). Here, it becomes evident that the MgS batteries are exclusively made of abundant materials, which minimizes the MDP impact to the extent that it becomes hardly visible in comparison to the battery pack periphery, such as pack housing and BMS. In contrast, except for the LFP (M-B) with the anode as the main contributor to MDP (62%), the lithium batteries show significantly higher MDP impacts. They are driven mainly by the cathode (53% NMC (Ell); 62% NMC (M-B); 47% LFP (Zak); 44% LiS (Deng)), and the anode materials (24%, 26% and 31%, 40% of the total for NMC (M-B), NMC (Ell), LiS (Deng), and LFP (Zak), respectively). For the anode, the main reason for this is the copper current collectors. In contrast, for the cathode, the main reason is the use of cobalt and nickel in the cathode active material (NMC) and the aluminium current collector (LiS).

### Sensitivity analysis

Electricity mix

The following sensitivity analysis assesses the influence of the electricity mix within our battery model. As the MgS-Evo2 exhibits the best environmental performance, only this cell configuration is considered.

The electricity supply for cell and battery pack manufacture is the central hot spot identified for the environmental impacts of the MgS-Evo2. Thus, the use of the Chinese and Swiss electricity mixes, the representative for an electricity mix of high (CN) and low carbon intensity (CH), is considered (for the base case, the average EU electricity mix was assumed), with the corresponding results given in Figure 5. The composition of the EU and Swiss electricity mixes respectively include 30% and 33% of nuclear power, in contrast with 2% in the Chinese electricity mix[35]. As previously mentioned, due to the lack of information about possible commercial manufacturing lines for MgS batteries, we initially assumed that MgS and LIB manufacturing processes would have comparable energy demands. This assumption might turn out to be inaccurate since one of the key drivers for manufacturing energy demand is the dry room required for handling the highly hygroscopic electrolyte of LIB. The MgS electrolyte used for the prototype cell (Mg[B(hfip)$_4$]$_2$) (0.3M) is much less hygroscopic, and an optimized production line might show significantly lower energy demand. However, in terms of environmental impacts, until information about the actual manufacturing process of MgS batteries on a commercial level becomes available, this might have similar consequences in the four assessed impact categories as changing to a low-carbon electricity mix.

Magnesium anode foil thickness

Being the anode foil one of the drivers of environmental impacts in several of the assessed categories, we optimise the Mg foil thickness (in the prototype pouch cell determined by the simple availability of Mg foil with given thickness on the market) according to electrochemical considerations. For this purpose, we calculate the area of the Mg foil based on the information available for the pouch cell, and the mass of aluminium required to match the electrochemical potential of the sulfur cathode. Being the Mg foil both active material and current collector, a minimum thickness needs to be maintained also in fully discharged state in order to assure current collection and avoid increasing ohmic losses. The minimum foil thickness for this purpose is estimated based on the conductivity of aluminium and magnesium, using the aluminium cathode thickness as reference, and increasing the Mg foil thickness proportional to its higher specific ohmic resistance. Being the conductivity of magnesium roughly half that of aluminium, and the thickness of the cathode collector foil 4.4um (88mg for an area of 74 cm2 in the prototype pouch cell), the resulting minimum magnesium foil thickness in discharged state is 8.8 um, equivalent to 113 mg Mg per pouch cell. The additional amount of magnesium required to match the specific capacity of Sulfur (1.67Ah/g) can be estimated based on electrochemical considerations. With a specific capacity of sulfur of 2.21 Ah/g and a sulfur cathode load of 421mg/cell (see Table 1), the required amount of Magnesium anode active material amounts to 319 mg/cell. Adding the estimated minimum foil thickness for assuring current collection functions even in fully discharged state (113 mg), we obtain an optimized Mg foil mass of 432 mg, already slightly above that of the pouch cell prototype. We can therefore state that, even if not directly intended but rather driven by pragmatic reasons, the magnesium foil thickness is already at the lower limit and does not hold any further optimization potential.

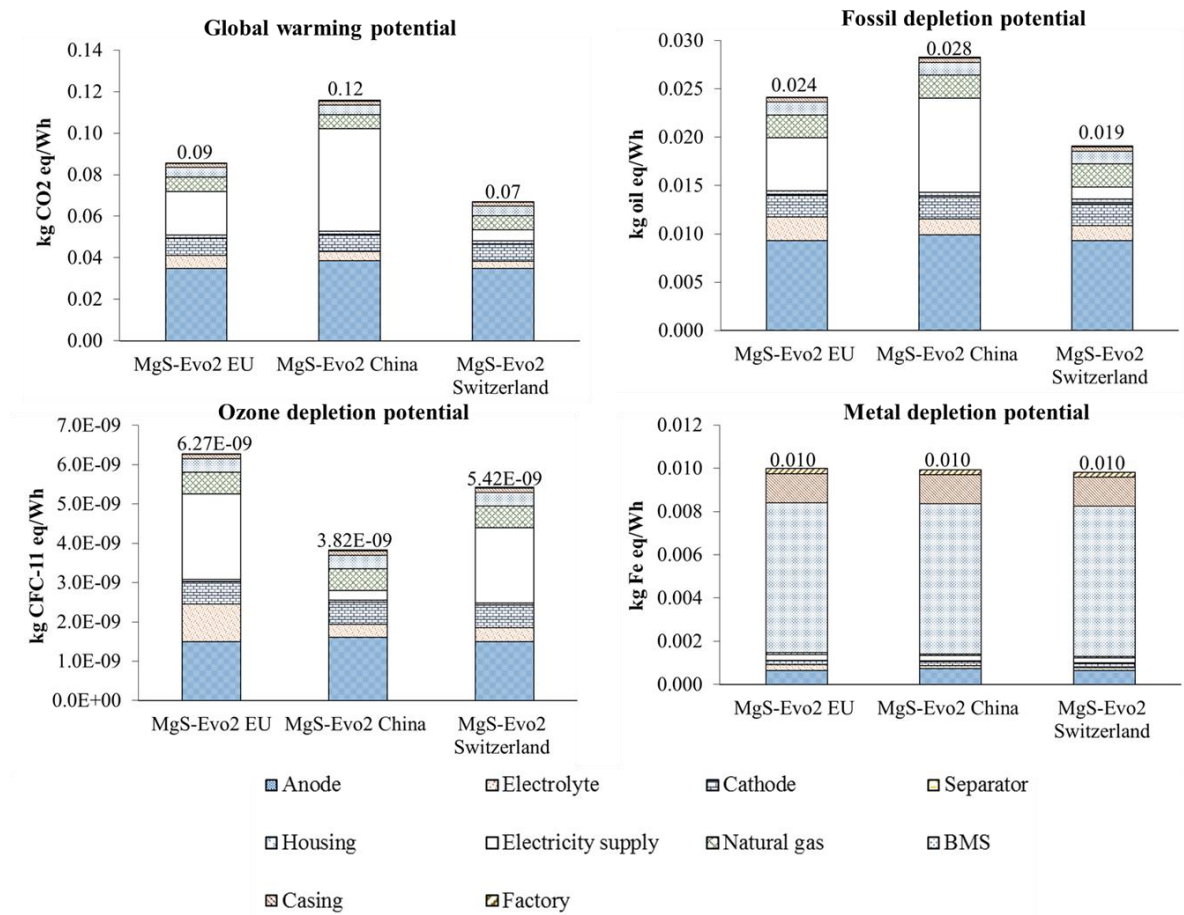

**Figure 5** Effect of the electricity mix on the environmental impacts associated with MgS-Evo2. Legend: MgS-Evo2 EU, MgS-Evo2 China, MgS-Evo2 Switzerland, MgS-Evo2: second evolution with optimized separator thickness and the correspondingly lower amount of electrolyte, respectively considering the European, Chinese and Swiss electricity mix.

## 4 Discussion

The results show that once the configuration of the MgS cell is optimized regarding the pouch cell housing and the separator (MgS-Evo2), the composition of the electricity mix defines the pathway to reduce its environmental impacts further. As discussed above, apart from the cell manufacturing process itself, attention should also be given to the anode manufacturing process, which contributes the second significant share and quickly becomes dominant when the impact from electricity for cell manufacture decreases. Interestingly, the electricity supply in the magnesium manufacturing process is the critical source of GHG emissions, so even here the sourcing of magnesium obtained by using 'green' electricity would be the clue to the improvement of environmental performance. Efforts to improve the carbon footprint of the MgS battery should, therefore, focus on reducing the environmental impacts of magnesium production (magnesium metal production is highly energy-intensive, and the corresponding energy mix will be essential for this purpose). The effect of the optimizations assumed for the MgS cells is also evident, eliminating the key contributors to GWP, including the large and heavy cell pouch of the prototype MgS-BL, and subsequently optimizing the thickness of the separator and thus the amount of electrolyte (MgS-Evo2).

While the results for FDP are similar to those of GWP, the ODP results show different behaviour. Here, the Chinese (coal-based) electricity mix gives better results; this is explained by the emissions from uranium and natural gas manufacture, which overall are dominant for the CFC-114 and Halon-1301 emissions to air. These emissions from Swiss and European electricity mixes contribute 35% and 29%, and 30% and 26% respectively to the total estimated ODP. Regarding metal depletion, cell manufacturing energy plays an insignificant role in this impact category. Conversely, the composition of the electricity mix used for cell manufacture or variations in the amount of electricity is negligible (see Figure 5).

When comparing the $CED_{total}$ of the MgS batteries with the corresponding average value reported for LIB (1.81 MJ/Wh) [28], the MgS-Evo2 achieves a lower cumulative energy demand. The main contributor to the total CED for this optimized cell configuration is electricity supply (34% of the total; mainly the energy requirements associated with cell manufacture), indicating this as a critical assumption in the present assessment. Apart from electricity, anode manufacture is the second contributor to MgS-Evo2, and here again the electricity mix (and its carbon intensity) used for producing the magnesium metal is the key for further reducing environmental impacts. A reduction of the content of magnesium metal within the battery cell (as possible further evolution or optimization step) is found to be not reasonable, being the magnesium foil thickness already at the lower limit for maintaining both electrochemical and current collector functions. In contrast, for the MgS-Evo1 cell configuration, the (non-optimized and therefore high amount of) electrolyte plays a significant role, and in the case of MgS-BL, the cell pouch contributes substantially.

In summary, the optimized MgS batteries obtain promising results for all four evaluated impact categories. In this first evolution stage (where only a reduction of the cell pouch mass to values similar to those of current LIB is assumed; MgS-Evo1), their environmental impacts are already similar to or below the LiS battery. If the second evolution stage can be achieved MgS batteries have the potential to outperform also current LIB, regarding the production phase.

# 5 Conclusions

We assessed the environmental performance of an MgS battery in three different configurations; a prototype cell based on actual data from a project, and two hypothetical evolutions of this, with a theoretical optimization of the cell layout according to the current state of the art in lithium-ion battery technology. The first prototype cell shows a comparably poor environmental performance due to massive mass of pouch housing, thick separator and high amount of electrolyte. However, the optimized cell layouts show a very promising performance, with GWP impacts comparable to or even better than those of current NMC and LFP type (LIB), and low impacts in mineral (metal) resource depletion. A remaining key contributor to the majority of the four assessed environmental impact categories is electricity demand along the process chain, in particular cell manufacture, where a similar energy intensity to current LIB is assumed. Moreover, since the electrolyte of the MgS battery is less hygroscopic, potentially avoiding the need for a dry room, which is one of the key drivers of manufacturing energy demand for LIB, further reduction in the associated environmental impacts of the MgS cell can be expected. If the magnesium production process were to rely on renewable electricity, this would further reduce its environmental impacts.

If the assumed improvements to the current MgS cell layout can be achieved, then the MgS battery has a high potential for outperforming its competitors LIB and LiS under environmental aspects. However, this will only be possible if parameters that can be equally important for the total environmental performance can also be achieved, e.g. comparable technical performance in terms of efficiency and lifetime. These are not yet foreseeable and so are not evaluated further here, but they remain highly relevant for future research regarding the overall environmental impacts of MgS batteries.

## Acknowledgments

This work contributes to the research performed at CELEST (Center for Electrochemical Energy Storage Ulm-Karlsruhe) and was funded by the German Research Foundation (DFG) under Project ID 390874152 (POLiS Cluster of Excellence), and was financially supported by the Initiative and Networking Fund of the Helmholtz Association within the Network of Excellence on post-Lithium batteries (ExNet-003). Jens Peters acknowledges funding from the European Union's Horizon 2020 research and innovation programme under the Marie Skłodowska-Curie grant agreement No 754382."